\documentclass[11pt,tightenlines,eqsecnum,floats,aps,amsmath,amssymb,nofootinbib,prd,shownopacs]{revtex4}
\usepackage{amsmath,amssymb,amsfonts,amsthm}
\usepackage{graphicx}
\usepackage{enumerate} % advanced enumerate environment
\usepackage{colordvi} % for color text
\usepackage[mathscr]{eucal}
\usepackage[latin1]{inputenc}
\usepackage{amsmath}
\usepackage{amsfonts}
\usepackage{amssymb}
\usepackage{graphicx}

\def\lp{{\ell}_{\rm Pl}}

\newcommand{\f}{\frac}

\def\f{\frac}

\def\d{\textrm{d}}

\newcommand{\be}{\nopagebreak[3]\begin{equation}}
\newcommand{\ee}{\end{equation}}
\newcommand{\ba}{\nopagebreak[3]\begin{eqnarray}}
\newcommand{\ea}{\end{eqnarray}}

\newcommand{\bmult}{\nopagebreak[3]\begin{multline}}
\newcommand{\emult}{\end{multline}}

\def\d{{\rm d}}

\def\lp{{\ell}_{\rm Pl}}

\def\f{\frac}

\def\d{\textrm{d}}

\def\lp{l_{\rm Pl}}

\def\d{{\rm d}}

\usepackage{enumerate}
\begin{document}

%\markboth{Parampreet Singh}
%{Is classical flat Kasner spacetime flat in quantum gravity?}

%\catchline{}{}{}{}{}
%
%%%%%%%%%%%%%%%%%%%%%%%%%%%%%%%%%%%%%%%%%%%%%%%%%%%%%%%%%%%%%%%%%%%%
\title{Is classical flat Kasner spacetime flat in quantum gravity?}

%\author{PARAMPREET SINGH}

\author{Parampreet Singh}
\affiliation{Department of Physics and Astronomy,\\
Louisiana State University, \\ Baton Rouge, LA 70803, U.S.A.}

%\address{{Department of Physics and Astronomy, Louisiana State University, Baton Rouge, LA 70803, USA}\\
%psingh@phys.lsu.edu}

%\maketitle

%\begin{history}
%\received{Day Month Year}
%\revised{Day Month Year}
%\end{history}

\begin{abstract}
 
Quantum nature of classical flat Kasner spacetime is studied using effective spacetime description in loop quantum cosmology.
We find that even though 
the spacetime curvature vanishes at the classical level,  non-trivial quantum gravitational effects can arise. 
For the standard 
loop quantization of Bianchi-I spacetime, which uniquely yields universal bounds on expansion and shear scalars and results in a generic 
resolution of strong singularities, we find that a flat Kasner metric is not a physical solution of the effective spacetime description, 
except in a limit. The lack of a flat Kasner metric at the quantum level results from a novel feature of the loop quantum 
Bianchi-I spacetime: quantum geometry induces non-vanishing spacetime curvature components, making it not Ricci flat even when no matter is 
present. The non-curvature singularity of the classical flat Kasner spacetime is avoided, and 
the effective spacetime transits from a flat Kasner spacetime in asymptotic future, to a Minkowski spacetime in asymptotic past.
Interestingly, for an alternate loop quantization which does not share some of the fine features of the standard quantization, flat Kasner 
spacetime with expected classical features exists. In this case, even with non-trivial quantum geometric effects, the spacetime curvature 
vanishes. These examples show that the character of even a flat classical vacuum spacetime can alter in a fundamental way in quantum gravity and is 
sensitive to the quantization procedure.

\end{abstract}
\maketitle
%\keywords{Loop quantum cosmology, Anisotropic spacetimes, Quantum gravity }

%\ccode{PACS numbers:}

%\tableofcontents

\section{Introduction}	
An important issue in quantum gravity is to understand the way fundamental properties of 
classical spacetime change when quantum gravitational effects become important.
For classical spacetimes where dynamics of gravitational field causes curvature singularities, it has been 
expected that quantum gravity effects resolve the latter. Recent studies in loop quantization of cosmological and black hole spacetimes indeed 
confirm these expectations. Nonetheless, there exist classical spacetimes where curvature singularities are absent. One such spacetime is the flat Kasner spacetime which is a special solution of  the vacuum Bianchi-I spacetime in general relativity. It is obtained from the Kasner metric when one of the Kasner exponents is unity, and the other two vanish. The flat Kasner spacetime   
has vanishing components of the Riemann curvature and a non-curvature singularity at the origin. It can be embedded in Minkowski space. At 
first, it may appear that such a spacetime with an innocuous gravitational field probably is too simple for any non-trivial quantum 
gravitational effects. But, this expectation 
may be premature. 
%After all, the spcaetime is just a patch of flat space. 
Even though Riemann curvature vanishes, one of the directional Hubble rates diverges, causing expansion and the shear scalars to blow up at 
the non-curvature singularity. Therefore, it is not obvious whether quantum gravity does not affect classical properties of the flat Kasner 
spacetime.

The goal of this manuscript is to understand quantum properties of classical flat Kasner spacetime in loop quantum cosmology (LQC), which is a non-perturbative 
quantization of homogeneous spacetimes \cite{as}. LQC is based on using techniques of loop quantum gravity where elementary gravitational phase space variables are the triads and the conjugate connection components. At the quantum level, holonomies of connection are used to construct a curvature operator whose non-local nature results in a quantum Hamiltonian constraint which turns out to be a quantum difference equation. In the last decade, loop quantization of various homogeneous gravitational spacetimes has been successfully performed. 
 One the main results is the absence of a big bang and an existence of a non-singular bounce \cite{aps,aps3,acs}, a direct result of the underlying quantum discreteness. Using consistent histories analysis, the probability of bounce turns out to be unity \cite{consistent}. Resolution of singularities in LQC can be understood quite accurately using an effective spacetime description obtained using geometric formulation of quantum theory \cite{vt}. The effective Hamiltonian in LQC can also be obtained using an inverse procedure based on deriving canonical structures from the dynamical equations. Demanding that the modified dynamical equations resolve the singularity through quadratic repulsive energy density, as in LQC, one is led to a canonical phase space and the Hamiltonian of LQC \cite{ss15}. The effective spacetime description turns out to be an excellent approximation to the underlying quantum dynamics \cite{aps3,khanna,ps12,dgs2,squeezed}. Using effective dynamics, various insights on the quantum 
geometric 
effects have been gained. In particular, quantum geometry causes universal bounds on directional Hubble rates, and hence on  
 expansion and shear scalars \cite{cs09,ps09,js1}. It results in a generic resolution of strong singularities for isotropic and anisotropic spacetimes (at least for matter with a vanishing anisotropic stress) \cite{ps09,psvt,ps11,ps15,ss1}. 
 
 It should be noted that 
 different ways to consider holonomies over loops lead to different loop quantization prescriptions. These result in different quantum Hamiltonian constraints with qualitatively different physics. In particular,  
  existence of universal bounds on Hubble rates, expansion and shear scalars, and a generic resolution of strong singularities turn out to 
be true only for a unique prescription in LQC \cite{cs08,cs09}: the improved dynamics quantization \cite{aps3}. This quantization method has 
an advantage of being independent of the freedoms of the fiducial cell necessary to define the  symplectic structure.  Apart from the 
improved dynamics quantization of the Bianchi-I spacetime \cite{cv,awe},  denoted as $\bar \mu$ scheme in our analysis, there exists an 
alternative loop quantization  \cite{chiou,b1madrid1,b1madrid2,szulc_b1} (denoted as $\bar \mu'$ scheme).\footnote{In earlier works on 
Bianchi-I models \cite{chiou,cv}, notation of $\bar \mu$ and $\bar \mu'$ were reversed.} Unlike the (standard) $\bar \mu$ 
quantization, the $\bar \mu'$ quantization does not yield consistent physics for the case of a $\mathbb{R}^3$ spatial manifold. Physics 
depends on the change in the shape of the fiducial cell defined on $\mathbb{R}^3$. However, if 
one chooses,
 a 3-torus spatial manifold then this issue is 
overcome. In any case, directional Hubble rates are not universally bound and resolution of strong singularities is not guaranteed in $\bar \mu'$ scheme \cite{cs09,ps11,ps15}. 
  %Since it shares all the features of the improved dynamics in the isotropic case, the  $\bar \mu$ scheme is considered as the standard loop quantization of the Bianchi-I spacetime. 
  Nevertheless it is a reasonable alternative quantization to study if the spatial manifold is restricted to a 3-torus.

 Our investigations of the flat Kasner spacetime in LQC are based on both of the above quantization prescriptions. It utilizes the effective spacetime description, which for the Bianchi-I spacetime has been well tested using numerical simulations of physical states with the quantum Hamiltonian constraint \cite{b1madrid2,djms}. Our analysis is based on assuming the validity of the effective Hamiltonian for the entire regime. %Some surprising results are found in our analysis. 
 With this caveat, the summary of the results we obtain from effective Hamiltonian is as follows.  
 For the  effective dynamics   of the standard ($\bar \mu$) loop  quantization of the Bianchi-I spacetime in the absence of matter, we find that flat Kasner spacetime is not a solution. Only in the 
 classical limit at late times, the effective spacetime approaches the flat Kasner spacetime. Unlike the classical theory, the effective spacetime has non-vanishing Riemann tensor components, and the spacetime is not Ricci flat. 
 Departure from the classical properties -- flat spacetime and vacuum, is caused by the underlying quantum geometric discreteness. 
 
 Evolution of directional scale factors reveals some more surprises in the $\bar \mu$ quantization. At late times when spacetime curvature 
is negligible in the effective spacetime, two of the scale factors remain constant and the third varies linearly with proper time, as 
should be the case for the flat 
Kasner metric. There is an excellent agreement with classical theory in this regime. In the backward evolution, spacetime curvature 
components grow rapidly near the origin, the point of non-curvature singularity in the classical theory. As the spacetime curvature in the 
effective spacetime becomes Planckian, the linearly varying directional scale factor changes its behavior and avoids the origin. In the meantime, other scale factors remain constant.  Curvature invariants are bounded in this regime due to quantum 
geometry and evolution mimics the behavior as if a to be curvature singularity is resolved. In the past evolution, past to the avoidance of the point where 
classical non-curvature singularity exists, the spacetime again 
becomes classical. But, it no longer  approaches a flat 
Kasner spacetime 
asymptotically. Interestingly, in this regime  all  of the three scale factors become constant. The effective spacetime at very early times approaches a Minkowski spacetime. 
 
 In the alternative ($\bar \mu'$) approach, evolution is very distinct from the $\bar \mu$ scheme.  It turns out that even in the presence of non-trivial quantum geometric effects, evolution  mimics the one in the classical theory. 
 As in general relativity, one of the directional Hubble rates is inversely proportional to time,  and other two directional Hubble rates vanish for all times. All the components of the spacetime curvature vanish.
Quantum geometry ignores the non-curvature singularity at the origin. This behavior is similar to ignorance of weak singularities by 
effective dynamics in LQC \cite{ps09,psvt,ps11,ps15,ss1}. 

Though, evolution of scale factors is essentially the same as in the classical theory for the $\bar \mu'$ quantization, quantum geometry does leave its subtle trace. It renormalizes a coefficient in the directional Hubble rate, from unity in the classical theory to another finite value. 
Metric of the effective spacetime turns out to be a flat Kasner metric. In a sharp contrast to the standard $(\bar \mu)$ loop quantization, the classical flat vacuum spacetime remains a flat vacuum spacetime at the quantum level. Thus, there are striking qualitative differences in the effective dynamics of the standard loop quantization and its alternative in the Bianchi-I spacetime. 
%On one hand, the standard quantization yields a non Ricci-flat spacetime purely due to quantum discreteness 

Analysis and result in the manuscript are organized as follows. In the next section, we summarize the Hamiltonian approach using connection and triad variables for the classical Bianchi-I spacetime in the absence of matter. 
By studying the solutions of the Hamilton's equations, flat Kasner metric is shown to be a solution in the classical theory. In Sec. III, effective Hamiltonian dynamics of $\bar \mu$ quantization is 
considered with no matter. Unlike the classical case, it is found that the flat Kasner metric is no longer a solution of the effective Hamiltonian. Using the modified Hamilton's equations, the effective spacetime is shown to be not Ricci flat. 
Curvature invariants are found, and shown to be non-vanishing but bounded. Underlying quantum geometry thus changes the fundamental characteristics of the classical flat vacuum spacetime. Sec. IV deals with a similar analysis of an alternative loop quantization, the $\bar \mu'$ scheme. In this case, the effective spacetime is Ricci flat, and flat Kasner metric turns 
out to be a solution of the effective Hamiltonian. Unlike the $\bar \mu$ effective dynamics, the non-curvature singularity at the origin is not avoided. We conclude with a discussion  in Sec. V.

\section{Classical vacuum Bianchi-I spacetime and the flat Kasner solution}
%In this section, we summarize ...
The gravitational phase space of the homogeneous Bianchi-I spacetime is composed of triad and connection variables, which due to underlying symmetries of homogeneity are diagonal and only depend on time. These satisfy,
\be
\{c_i,p_j\} = 8 \pi G \gamma \delta_{ij} ~.
\ee
Here $\gamma$ denotes the Barbero-Immirzi parameter. The spacetime metric for the Bianchi-I spacetime is given by 
 \be
\d s^2 = g_{\mu \nu} \d x^\mu \d x^\nu = - \d t^2 + a_1^2(t) \d x_1^2 + a_2^2(t) \d x_2^2 + a_3^2(t) \d x_3^2 ~,
\ee
where we have chosen the lapse to be unity, and $a_i$ denote the directional scale factors. These scale factors are kinematically related to the components of the 
triads, as follows: %Their time derivatives are related to the connection components. 
%The relationship between scale factors and triads is:
\be\label{triad-scalar}
p_1 = l_2 l_3 \, a_2 a_3, ~~~ p_2 = l_3 l_1 \, a_3 a_1, ~~~ \mathrm{and}~~~ p_3 =l_1 l_2 \,  a_1 a_2 ~.
\ee
Here $l_i$ refer to the fiducial lengths of a fiducial cell ${\cal V}$, on the spatial manifold $\mathbb{R}^3$, introduced to define a symplectic structure. 
The connection components are related to the time derivatives of the scale factors, a relationship which is determined by the 
%The relationship between the connection components and the time derivatives of the scale factor are determined from the 
Hamilton's equations. 

In this homogeneous setting, gauge and spatial diffeomorphism constraints are fixed and the only non-trivial constraint is the Hamiltonian constraint. 
In the absence of matter, it is given by
\be\label{hamcl}
{\cal C}_{\rm{cl}} = -\frac{1}{8 \pi G \gamma^2 V} (c_1 p_1 c_2 p_2 + c_2 p_2 c_3 p_3 + c_3 p_3 c_1 p_1)  \approx 0
\ee
where $V = (p_1 p_2 p_3)^{1/2}$ denotes the physical volume of the cell ${\cal V}$. Using the Hamiltonian, the Hamilton's equations for $p_i$ yield:
\be\label{dotpicl}
\dot p_i = \frac{p_i}{\gamma V} \left(c_j p_j + c_k p_k\right)  
\ee
with $i,j,k = 1..3$, which take different values in the above equation. Similarly, the Hamilton's equation for connection components turn out to be,
\be\label{dotcicl}
\dot c_i = -\frac{c_i}{\gamma V} \left(c_j p_j + c_k p_k\right) ~. 
\ee
%with similar equations for $c_2$ and $c_3$. 

Using eqs.(\ref{triad-scalar}), the directional Hubble rates are related to the time variation of triads as 
\be\label{Hidef}
H_i = \frac{\dot a_i}{a_i} = \frac{1}{2} \left(\frac{\dot p_j}{p_j} + \frac{\dot p_k}{p_k} - \frac{\dot p_i}{p_i} \right) ~.
\ee
%with $i,j,k = 1..3$ take different values. 
For the classical theory, this equation yields a simple relation between the directional Hubble rates and the connection components on using the Hamilton's equations (\ref{dotpicl}) and (\ref{dotcicl}):
\be
 H_i = \frac{c_i p_i}{\gamma V} 
\ee 
 which implies $c_i = \gamma l_i \dot a_i$. As we will see, this relationship changes in LQC.\\
%\ee
%which implies $c_i = \gamma l_i \dot a_i$.

The directional Hubble rates can be used to define two useful scalars, the expansion and the shear scalars. The expansion scalar $\theta$ is defined as
\be\label{exp-scalar}
\theta = \frac{\dot V}{V} = H_1 + H_2 + H_3 ~,
\ee
and the shear scalar is given by
\be
\sigma^2 = \frac{1}{3} \left((H_1 - H_2)^2 + (H_2 - H_3)^2 + (H_3 - H_1)^2\right) ~.
\ee

It is straightforward to verify using the Hamilton's equations that for the classical vacuum Bianchi-I spacetime, the time derivative of the directional 
Hubble rates satisfy
\be\label{Hieq}
\dot H_i + \theta H_i = 0 ~.
\ee
As a consequence, the time derivative of the expansion scalar obeys,
\be\label{thetaeq}
\dot \theta + \theta^2 = 0 ~.
\ee
 Using the relationship between the components of the Ricci tensor and the directional Hubble rates and expansion scalar, one can verify that eq.(\ref{Hieq}) implies $R^1_{1} = R^2_{2} = R^3_{3} = 0$. The vanishing of the classical Hamiltonian (eq.(\ref{hamcl})) which yields 
\be\label{hicons}
H_1 H_2 + H_2 H_3 + H_3 H_1 = 0,
\ee
when used with eq.(\ref{thetaeq}), implies that $R^0_{0} = 0$. Note that  we have obtained eqs.(\ref{Hieq}) and 
(\ref{thetaeq}) from the Hamilton's equations 
corresponding to the classical vacuum Bianchi-I Hamiltonian. One can alternatively obtain these equations, starting from the Einstein field equations and 
using the vacuum condition -- the vanishing of energy density and pressure, which leads to the vanishing of the Ricci tensor components.

Integrating (\ref{thetaeq}) and (\ref{Hieq}), we find 
\be \label{Hubble-Kasner}
H_1 = n_1 t^{-1}, ~~~ H_2 = n_2 t^{-1} ~~~ \mathrm{and} ~~~ H_3 = n_3 t^{-1} ~
\ee
where $n_i$ are constants. 
The definition of the expansion scalar (\ref{exp-scalar}) then yields:
\be\label{kasner1}
n_1 + n_2 + n_3 = 1 ~.%\f{gmail\sin(\bar \mu_3 c_3)}{\bar \mu_3}
\ee
Using eq.(\ref{hicons}), the square of the expansion scalar results in
\be\label{kasner2}
n_1^2 + n_2^2 + n_3^2 = 1 ~.
\ee
Using (\ref{Hubble-Kasner}), the spacetime metric can thus be written as the Kasner metric
\be
\d s^2 = -\d t^2 + t^{2 n_1} \d x_1^2 + t^{2 n_2} \d x_2^2 + t^{2 n_3}\d x_3^2 ~,
\ee
with $n_i$ referred to as the Kasner exponents.

In the case when one of the Kasner exponents is unity and other two vanish, all of the Riemann curvature components vanish. 
%The flat Kasner metric is obtained when either one of Kasner indices is unity, and others must vanish due to (\ref{kasner1}) and (\ref{kasner2}). 
Let us choose the non-vanishing exponent to be $n_1$. In this case, the metric becomes 
\be \label{flkasner}
\d s^2 = -\d t^2 + t^{2} \d x_1^2 + \d x_2^2 + \d x_3^2 ~
\ee
which can be transformed to a metric of a flat space using a coordinate transformation. Due to this reason, the Kasner metric with any of 
the Kasner exponents unity and other vanishing, is called the flat Kasner metric. The classical flat Kasner spacetime has a 
non-curvature singularity at $t=0$. Here the expansion and shear scalar diverge due to divergence in one of the directional Hubble rates 
(in the above example, $H_1$). Finally it is useful to note that the conditions 
for the flat Kasner metric (\ref{flkasner}), translate to $c_2 = c_3 = 0$ and $c_1$ as a non-vanishing constant. 
%choosing fiducial lengths $l_i$ to be unity, translate to the following conditions on the classical phase space variables:
%\be
%p_1 = 1, ~~~ p_2 = p_3 = t
%\ee
%and 
%\be
%c_1 = \gamma, ~~~ c_2 = c_3 = 0 ~.
%\ee
 Classical Hamilton's equations (\ref{dotcicl}) imply that if any of the two connection components vanish initially, then  all the 
connection  components are preserved in time. If the initial conditions are chosen such that $c_2 = c_3 = 0$, eq.(\ref{kasner1}) and 
(\ref{kasner2}) imply $c_1 = \gamma l_1$ at all times. Further, the Hamiltonian constraint is trivially satisfied for all times.

%The classical Hamiltonian constraint in terms of the connection and triads variables is

\section{Bianchi-I spacetime in LQC and the lack of flat Kasner solution}
Let us investigate the existence of flat Kasner metric in the standard ($\bar \mu$) loop quantization of Bianchi-I spacetime in LQC. 
Our analysis will be based on the 
effective Hamiltonian, which is assumed to be valid at all the scales. The effective Hamiltonian of the loop quantization of Bianchi-I spacetime was first considered 
by Chiou and Vandersloot \cite{cv,chiou}. The corresponding quantum theory was later analyzed by Ashtekar and Wilson-Ewing \cite{awe}. The effective dynamics of this 
quantization has been well studied \cite{cv,chiou,magnetic,cs09,ps11,corichi,gs1,gs2}. Some of the interesting results include: universal boundedness of expansion and shear scalars \cite{cs09}, resolution of all strong singularities for matter with a vanishing anisotropic stress \cite{ps11,ps15}, genericity of inflation after the bounce in the presence of anisotropies \cite{gs2} and Kasner transitions across the bounce \cite{gs1}. % (see also Ref. \cite{barrau}\footnote{However, note that Ref. \cite{barrau} not consider anisotropic shear as obtained from expansion tensor  .}
As discussed earlier, in literature other loop quantizations of this spacetime have also been considered. 
The property which distinguishes this quantization from other loop quantizations of Bianchi-I spacetime is that under rescalings of the 
fiducial lengths of the fiducial cell ${\cal V}$, physical predictions are not affected \cite{awe,cs09}. To understand this, one can 
consider the way trigonometric terms in the effective Hamiltonian (\ref{effham_bianchi}) transform under the change of fiducial lengths. It 
turns out that the variation in connection components cancels exactly with that in $\bar \mu_i$'s \cite{cs09}. As a result, there is a well 
defined ultra-violet scale at which the quantum effects become important, and an infra-red limit as general relativity at small spacetime 
curvatures. These features are in common to the improved dynamics loop quantization of the isotropic models \cite{cs08} and the 
Kantwoski-Sachs spacetimes \cite{js1}. %this loop quantization of the Bianchi-I model is singled out for having desired features of fiducial 

In the absence of matter, the effective Hamiltonian for the Bianchi-I spacetime in LQC is \cite{cv}%\footnote{In our notation, $\bar \mu$ which is a measure of holonomy length corresponds to $\bar \mu'$ in Ref. \cite{cv}.} :
\ba\label{effham_bianchi}
C_H^{\bar \mu} &=& \nonumber  - ~\f{1}{8 \pi G \gamma^2 V}\bigg[\f{\sin(\bar \mu_1 c_1)}{\bar \mu_1} \f{\sin(\bar \mu_2 c_2)}{\bar \mu_2}  p_1 p_2 
 + \f{\sin(\bar \mu_3 c_3)}{\bar \mu_3} \f{\sin(\bar \mu_1 c_1)}{\bar \mu_1} p_3 p_1 \\ && ~~~~~~~~~~~~~~~~~~~~ + \f{\sin(\bar \mu_2 c_2)}{\bar \mu_2} \f{\sin(\bar \mu_3 c_3)}{\bar \mu_3} p_2 p_3 \bigg] \approx 0 ~.
\ea
%\mathrm{cyclic} ~~ \mathrm{terms}\right) ~\\
%\ee
Here $\bar \mu_i$ are determined from the underlying quantum geometry. In particular, they are related to the minimum area gap $\Delta = 
\lambda^2 = 4 \sqrt{3} \pi \gamma \lp^2$ as follows \cite{awe}:
\be \label{mub1}
\bar \mu_1 = \lambda \sqrt{\f{ p_1 }{p_2 p_3}}, ~~~ \bar \mu_2 = \lambda \sqrt{\f{p_2}{p_1 p_3}}, ~~~ \mathrm{and} ~\bar \mu_3 = \lambda \sqrt{\f{p_3}{p_1 p_2}} ~ .
\ee
In the regime where $\bar \mu_i c_i \ll 1$, the effective Hamiltonian constraint approximates the classical Hamiltonian constraint 
(\ref{hamcl}). Since $c_i = \gamma l_i \dot a_i$ in the classical theory, in this regime $H_i \ll \lambda_i^{-1}$. Thus, when Hubble rate 
is much smaller than the inverse of Planck length we expect the effective dynamics obtained from the above Hamiltonian constraint to 
approximate the classical dynamics. 

However, in the regime where  Planck scale effects are important there are significant departures between the effective  dynamics obtained from (\ref{effham_bianchi}) and the classical dynamics. 
This is evident from the Hamilton's equations corresponding to the effective Hamiltonian in eq.(\ref{effham_bianchi}). The equations for the triads are:
\be
\dot p_i = \frac{p_i}{\gamma \lambda} \cos(\bar \mu_i c_i) \left(\sin(\bar \mu_j c_j) + \sin(\bar \mu_k c_k)\right) ~,
\ee
where $i,j,k$ take values $1..3$ independently. Similarly, the time variation of connection components is given by
\ba\label{ci2}
\dot c_i &=& \nonumber \frac{1}{2 p_i \gamma \lambda} \bigg[c_j p_j \cos(\bar \mu_j c_j) (\sin(\bar \mu_i c_i) + \sin(\bar \mu_k c_k)) + c_k p_k \cos(\bar \mu_k c_k) (\sin(\bar \mu_i c_i) + \sin(\bar \mu_j c_j)) \\ && ~~~~~~~~~~~~ - c_i p_i \cos(\bar \mu_i c_i) (\sin(\bar \mu_j c_j) + \sin(\bar \mu_k c_k)) - \bar \mu_i p_j p_k \bigg[\sin(\bar \mu_i c_i) \sin(\bar \mu_j c_j)\\ && ~~~~~~~~~~~~~~~~~~~~~~~~+ \sin(\bar \mu_j c_j) \sin(\bar \mu_k c_k) + \sin(\bar \mu_k c_k) \sin(\bar \mu_i c_i) \bigg]\bigg] ~.
\ea
The departure between the effective and the classical dynamics is notably captured by the behavior of the directional Hubble rates:
%The directional Hubble rates turn out to be
\be\label{Hi2}
H_i = \frac{1}{2 \gamma \lambda} \left(\sin(\bar \mu_j c_j + \bar \mu_k c_k) + \sin(\bar \mu_i c_i - \bar \mu_k c_k) + \sin(\bar \mu_i c_i - 
\bar \mu_j c_j\right) ~.
\ee
In contrast to the classical theory, $H_i$ are universally bounded.  An immediate consequence of which is that the expansion and shear scalars 
are also bounded, with universal maximum values given by \cite{cs09,gs0}
\be
\theta_{\rm max} = \f{3}{2 \gamma \lambda}, ~~~~ {\mathrm{and}} ~~~
{\sigma^2}_{\rm max}=\frac{10.125}{3 \gamma^2 \lambda^2} ~.
\ee
This feature plays an important role in resolution of strong singularities in the effective spacetime description of this loop quantized 
Bianchi-I model, and geodesic completeness \cite{ps09}. For various types of matter contents, the classical big bang singularity is 
resolved and replaced by a bounce of the directional scale factors in the Planck regime.

%It is useful to note some pecilar features of the behavior of Hubble rates i

It is useful to note some features of the behavior of Hubble rates in effective dynamics of LQC to understand the way they 
affect the Ricci tensor components and existence of (flat) Kasner metric. In effective dynamics $H_i \neq (\gamma l_i a_i)^{-1} c_i$ except 
in the classical regime $H_i \ll \lambda_i^{-1}$. %Thus, unlike in the classical theory the 
 Its consequence is that the  vanishing of the effective Hamiltonian constraint does not in general implies that directional Hubble rates satisfy (\ref{hicons}), unless for some special choice of $H_i$'s. One such case is when two of the directional Hubble rates vanish which is required for the flat Kasner metric in the classical theory. As we saw in Sec. II, for the Bianchi-I vacuum spacetime in general relativity,
 eq.(\ref{hicons}) follows from the Hamiltonian constraint. It holds for all physical solutions  without any further condition on the classical directional Hubble rates. % and the Ricci tensor component $R_{tt}$ turns out to be $R_{tt} = \dot \theta + \theta^2$. 
 Vanishing of $\dot \theta + \theta^2$ which is necessary to obtain Kasner metric with $n_1 + n_2 + n_3 = 1$, then implies $R^0_{0} = 0$. 
Unlike the classical theory, in LQC $R^0_0 \neq \dot \theta + \theta^2$ in general. %We reach a striking conclusion.
 Unless Hubble rates satisfy an additional constraint, existence of Kasner metric is not tied to the vanishing 
Ricci tensor components. 
 
 In the case when two of the directional Hubble rates, say $H_2$ and $H_3$, vanish for all times then $R^0_0 = \dot \theta + \theta^2$ and $R^1_1 = \dot H_1 + \theta H_1$. Both of these components equal each other, and other components of Ricci tensor vanish in LQC. The flat Kasner metric then exists if 
$\dot H_1 + H_1^2 = 0$. %xTo find the flat Kasner metric solution, it is important to understand the time variation of all the directional Hubble rates at all times in effective dynamics.

Let us now investigate in detail whether the flat Kasner metric is a solution  permitted by the effective Hamiltonian 
(\ref{effham_bianchi}). If such a metric exists, then directional Hubble rates 
should satisfy the time dependence $H_i = n_i t^{-1}$, with one of the $n_i$'s equal to unity and others vanishing. To determine whether 
such a temporal variation of Hubble rates is allowed throughout the evolution, let us choose the initial conditions on $c_2, c_3$ at time 
$t_i$ in the classical regime which agree with the conditions needed for the flat Kasner metric in the classical theory, i.e. $c_2(t_i) = 
c_3(t_i) = 0$. We then  seek whether the effective Hamiltonian constraint and the effective dynamics leads to the temporal evolution of phase 
space variables necessary for a flat metric. This choice of initial conditions implies that the directional Hubble rates $H_2$ and $H_3$ 
vanish at the initial time. Note that vanishing of $c_2$ and $c_3$ at initial time $t_i$ ensures that the effective Hamiltonian constraint is 
satisfied without any further constraints on any other phase space variables. Therefore, the initial conditions for directional Hubble rate 
$H_1$ are not fixed. 

Using the first order equations (\ref{ci2}), for the above initial conditions, we find that all of the connection components 
are constant in time. Hence, $c_2 = c_3 = 0$ at all times. Therefore, the effective Hamiltonian constraint is satisfied at all times 
without constraining other phase space variables.   For $c_2(t) = c_3(t) = 0$,
%not a coordinate singularity since the curvature invariants are non-vanishing. $, 
the time variation of triads yields
\be
\frac{\dot p_1}{p_1} = 0, ~~~ \frac{\dot p_2}{p_2} = \frac{1}{\gamma \lambda} \sin(\bar \mu_1 c_1) ~~~ \mathrm{and} ~~~ 
\frac{\dot p_3}{p_3} = \frac{1}{\gamma \lambda} \sin(\bar \mu_1 c_1) ~.
\ee
Using the above equations in eq.(\ref{Hidef}),  directional Hubble rates are given by 
\be
H_1 = \frac{1}{\gamma \lambda} \sin(\bar \mu_1 k),~~~~ H_2 = H_3 = 0 ~.
\ee
Here $k$ denotes the constant value of $c_1$ for a given fiducial cell. Since the directional Hubble rates, $H_2$ and $H_3$ vanish at all times, one 
finds that scale factors $a_2$ and $a_3$ are constant in time. Without any loss of generality, let us choose these scale factors to be unity to write 
%Choosing these scale factors to be unity,  we find 
\be
p_1 = 1, ~~~p_2 = p_3 = a_1 ,
\ee
where we have also chosen the fiducial lengths $l_i$ of the fiducial cell $\cal V$ to be unity. Above, the scale factor $a_1$ is an undetermined 
function of time.
\begin{figure}[tbh!]
\includegraphics[width=0.5\textwidth]{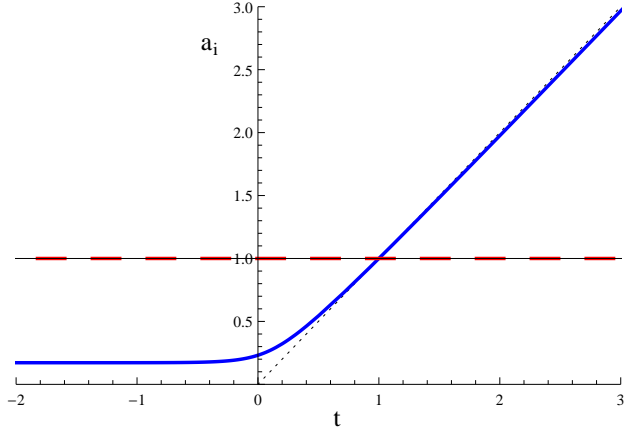}
%\centerline{\psfig{file=scalefactor.eps,width=8cm}}
%\vspace*{8pt}
\caption{The time variation of directional scale factors is shown for the effective Hamiltonian (\ref{effham_bianchi}) when two of the 
Hubble rates $H_2$ and $H_3$ vanish.  Scale factors $a_2$ and $a_3$ are shown by  dashed ticked and solid curves respectively. The scale 
factor $a_1$ is denoted by solid thick curve, and its classical counterpart is shown with a thin dashed curve. Classical curves for $a_2$ 
and $a_3$ are in agreement with those in LQC. The behavior of 
$a_1$ shows the lack of flat Kasner solution.}
\end{figure}

%\be
%\frac{\dot p_1}{p_1} = 0, ~~~ \frac{\dot p_2}{p_2} = \frac{1}{\gamma \lambda} \sin\left(\frac{\lambda c_1}{a_1}\right) ~~~ \mathrm{and} 
%~~~ \frac{\dot p_3}{p_3} = \frac{1}{\gamma \lambda} \sin\left(\frac{\lambda c_1}{a_1}\right) 
%\ee

It is now straightforward to see that the scale factor $a_1$ does not has the desired time variation for a flat Kasner metric. From the 
equation for the directional Hubble rate $H_1$ we obtain,
\be
\dot a_1 = \frac{a_1}{\gamma \lambda} \sin\left(\frac{\lambda k}{a_1}\right) ~.
\ee
In the regime where quantum geometric effects are negligible $a_1$ is linearly proportional to time as expected in the classical regime. However, when quantum geometric effects can not be ignored this proportionality fails. It can also be checked that $\dot H_1 + H_1^2 \neq 0$.  Hence, we conclude that for this effective 
spacetime description even though matter is absent, the flat Kasner metric is not a physical solution except in the classical limit.

Solving Hamilton's equations numerically reveals interesting features of the behavior of the scale factors in the effective dynamics which 
are shown in Fig. 1. The plot compares the evolution of different scale factors in the $\bar \mu$ quantization LQC and classical theory. The 
Hubble rates $H_2$ and $H_3$ are chosen as constant at initial time $t_i$ corresponding to late times. The scale factors $a_2$ and $a_3$ are 
found to 
remain constant through the evolution. Their evolution in LQC agrees with the 
classical evolution. At late times, the scale factor $a_1$ has a linear time dependence as expected in 
the classical theory. The classical scale factor $a_1$ vanishes at $t=0$ where a classical non-curvature  
singularity occurs. However, in LQC $a_1$ departs significantly from the classical theory in the Planck regime and avoids the point $t=0$. 
An interesting feature of the evolution is that in the further evolution, $a_1$ becomes constant and $H_1$ vanishes. At very early times, 
the effective spacetime metric approximates Minkowski metric. Thus, the Kasner exponent $n_1$ approaches unity asymptotically at late times 
($t \gg 0$), and vanishes asymptotically at very early times ($t \ll 0$).

\begin{figure}[tbh!]
\includegraphics[width=0.5\textwidth]{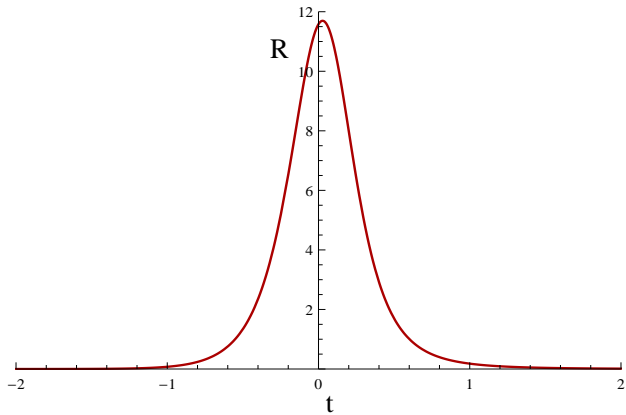}
%\centerline{\psfig{file=ricci.eps,width=8cm}}
%\vspace*{8pt}
\caption{Time variation of Ricci scalar for loop quantized Bianchi-I spacetime in absence of matter is shown in Planck units. Initial conditions correspond to those in Fig. 1. The Ricci scalar is non-vanishing in the absence of matter due to underlying quantum geometry.
 \label{f2}}
\end{figure}

To gain further insights on the nature of the effective spacetime, let us find the components of the Ricci tensor, and the curvature 
invariants. The Ricci tensor components turn out to be 
\be
R^0_{0} = R^1_{1} = \frac{1}{\gamma^2 \lambda} \sin\left(\frac{\lambda k}{a_1}\right) \left(\frac{1}{\lambda}\sin\left(\frac{\lambda k}{a_1}\right) - \frac{k}{a_1} \cos\left(\frac{\lambda k}{a_1}\right)\right)
\ee
and 
\be
R^2_{2} = R^3_{3} = 0 ~.
\ee
The striking result is that Ricci tensor components $R^0_0$ and $R^1_1$ are non-zero due to the underlying quantum geometry. {\it{The effective spacetime is not Ricci flat as in the classical theory in the absence of matter!}} 

The Ricci scalar $R$, which in this case is twice the value of $R^0_0$, is non-vanishing. Its time variation is shown in Fig. 2. The Ricci scalar approaches zero 
only in the classical regime $(H_i \ll \lambda_i^{-1})$. Notably the classical regimes have a different nature in the asymptotic past and 
asymptotic future. In the asymptotic future, 
where the Ricci scalar vanishes the effective spacetime metric approaches flat Kasner metric with one of the Kasner exponents unity and 
others zero. And in the asymptotic past, the vanishing of the Ricci scalar is tied to the purely Minkowskian metric with all Kasner 
exponents vanishing.

Like the Ricci scalar, the Kretschmann scalar ($K$) and the square of the Weyl curvature are also non-vanishing. These are respectively given by,
\be
K = \frac{1}{\gamma^4 \lambda^4 a_1^2} \bigg[\lambda k  \sin\left(\frac{2 \lambda k}{a_1}\right) - 2 a_1 \sin^2\left(\frac{\lambda k}{a_1}\right)\bigg]^2
\ee
and 
\be
C_{\alpha \beta \mu \nu} C^{\alpha \beta \mu \nu} = \frac{K}{3} ~. 
\ee
These share the same features as the Ricci scalar, and are vanishing in the asymptotic classical regimes. The nature of the spacetime near $t=0$ 
is thus very different in character than the non-curvature singularity in the classical theory. The singularity resolution 
resembles those of resolution of curvature singularities in LQC.

\section{Flat Kasner metric in an alternative Bianchi-I loop quantization}
In the previous section, we found that the standard $(\bar \mu)$ loop quantization of the Bianchi-I spacetime which 
results in a generic resolution of strong singularities does not admit a flat Kasner metric. Even in the absence of matter, Ricci tensor components became non-vanishing due to quantum geometric effects. A pertinent question is whether the absence of flat Kasner metric and non Ricci flatness is also a feature 
of other possible loop quantizations of Bianchi-I models. Investigation in this section is based on an alternative $(\bar \mu')$ loop quantization of Bianchi-I spacetime. 
The considered quantization was motivated by the improved dynamics quantization of the isotropic model, in the sense that the $\bar \mu_i'$ 
(used to construct holonomies) depend on the triads such that they seem a close generalization of their isotropic counterparts 
\cite{chiou}. 
 The particular dependence of $\bar \mu_i$ on the directional triads is the key difference from the loop quantization of the Bianchi-I 
spacetime studied in Sec. III. It should be noted that in both of the quantizations, $\bar \mu$ and $\bar \mu'$ prescriptions, agree with  
the isotropic model when the isotropic limit is taken. The main problem with this alternative loop quantization is the dependence of 
dynamics on the changes in the shape of the fiducial cell which is used to define the symplectic structure \cite{cs09}. As in the case of 
$\bar \mu$ prescription, this is straightforward to verify. The trigonometric terms in the effective Hamiltonian constraint 
(\ref{effham_bianchi_b}) are not invariant if the fiducial lengths $l_i$ are changed arbitrarily.  If the spatial manifold is a compact 
3-torus $\mathbb{T}^3$, then there is no need to introduce a fiducial cell and the problem disappears. Nevertheless, in this quantization 
the directional Hubble rate is not universally bounded, and in principle initial conditions allowing divergence in curvature invariants are 
possible. With these issues noted, we now investigate 
whether the effective dynamics of the alternative loop quantization of Bianchi-I model 
allows a flat Kasner metric.

%Let us consider a cpmpact 3-torus spatial manifold with radial directions of length $2 \pi$. 
The effective Hamiltonian for the alternative loop quantization of Bianchi-I spacetime in absence of matter is given by \cite{chiou,cv}:
\ba\label{effham_bianchi_b}
%\ba\label{effham_bianchi}
C_H^{\bar \mu'} &=& \nonumber  - ~\f{1}{8 \pi G \gamma^2 V}\bigg[\f{\sin(\bar \mu_1' c_1)}{\bar \mu_1'} \f{\sin(\bar \mu_2' c_2)}{\bar \mu_2'}  p_1 p_2 
 + \f{\sin(\bar \mu_3' c_3)}{\bar \mu_3'} \f{\sin(\bar \mu_1' c_1)}{\bar \mu_1'} p_3 p_1 \\ && ~~~~~~~~~~~~~~~~~~~~~~~~~~~~~~~~~~~~~~~ + \f{\sin(\bar \mu_2' c_2)}{\bar \mu_2'} \f{\sin(\bar \mu_3' c_3)}{\bar \mu_3'} p_2 p_3 \bigg] \approx 0 ~.
\ea
Here $\bar \mu_i'$, distinguished from the $\bar \mu_i$ in the previous section, are:
\be \label{mub1}
\bar \mu'_1 = \frac{\lambda}{\sqrt{{p_1}}}, ~~~ \bar \mu'_2 = \frac{\lambda}{\sqrt{p_2}}, ~~~ \mathrm{and} ~\bar \mu'_3 = \frac{\lambda}{\sqrt{{p_3}}} ~ ,
\ee
where $\lambda = (4 \sqrt{3} \pi \gamma \lp^2)^{1/2}$. Let us note a difference in comparison to the formulae in the previous section  
due to the compact 3-torus spatial manifold: the triads are related to the scale factors as $p_1 = (4 \pi)^2 a_2 a_3$, and similarly for 
$p_2$ and $p_3$.

Like in the case of the effective Hamiltonian in $\bar \mu$ quantization (\ref{effham_bianchi}), the above Hamiltonian constraint is 
approximated by the 
classical constraint when $\bar \mu_i' c_i \ll 0$. The Hamilton's equations for triads and connection components are:
\be
\frac{\dot p_i}{p_i} = \frac{1}{\gamma V} \cos(\bar \mu_i' c_i) \left(\frac{\sin(\bar\mu'_j c_j)}{\bar \mu_j'} p_j + \frac{\sin(\bar\mu'_k c_k)}{\bar \mu_k'} p_k \right)
\ee
and 
\be\label{dotci2}
\dot c_i = \frac{1}{2 \gamma V} \left(\frac{\sin(\bar \mu_j' c_j)}{\bar \mu_j'} p_j + \frac{\sin(\bar \mu_k' c_k)}{\bar \mu_k'} p_k \right) \left(c_i \cos(\bar \mu'_i c_i) - 3 \frac{\sin(\bar \mu_i' c_i)}{\bar \mu_i'} \right) ~.
\ee
In contrast to the effective dynamics of $\bar \mu$ quantization, $\dot p_i/p_i$ are not universally bounded. They can diverge if any of the scale factors vanish \cite{cs09}. The directional Hubble rates,
\ba
H_i &=& \nonumber \frac{1}{2 \gamma V} \bigg[p_i \frac{\sin(\bar \mu_i' c_i)}{\bar \mu_i'} (\cos(\bar \mu_j' c_j) + \cos(\bar\mu_k' 
c_k)) + p_j \frac{\sin(\bar \mu_j' c_j)}{\bar \mu_j'} (\cos(\bar \mu_k' c_k) - \cos(\bar\mu_i' c_i)) \\ &&~~~~~~~~~~~+ p_k \frac{\sin(\bar 
\mu_k' c_k)}{\bar \mu_k'} (\cos(\bar \mu_j' c_j) - \cos(\bar\mu_i' c_i)) \bigg] ~,
\ea
as a result are also not necessarily bounded. %Depending on the initial conditions, curvature invariants can even diverge. 
Due to these reasons, resolution of singularities is not guaranteed in this approach \cite{ps11,ps15}. Even in the cases where singularities are resolved and replaced by bounces of the scale factors, curvature invariants can take super-Planckian values in the bounce regime \cite{cs09}.

As in the $\bar \mu$ quantization,  physical solutions of the effective Hamiltonian constraint (\ref{effham_bianchi_b}) do not 
automatically satisfy eq.(\ref{hicons}). If directional 
Hubble rates satisfy the latter equation, then $\dot \theta + \theta^2$ is equivalent to $R^0_0$. Repeating the argument we made for the $\bar \mu$ quantization, the flat Kasner metric 
exists if for the non-vanishing directional Hubble rate, say $H_1$, $\dot H_1 + H_1^2 = 0$, with $H_2$ and $H_3$ vanishing at all times. 

To find whether flat Kasner metric is a solution in this effective spacetime description, we set the initial conditions as in the previous 
section, i.e.  the connection components  $c_2$ and $c_3$ both vanishing at an initial time $t_i$ in the classical 
regime. These ensure that the effective Hamiltonian constraint (\ref{effham_bianchi_b}) is satisfied initially, and, $H_2(t_i) = H_3(t_i) = 
0$. The first order differential equations (\ref{dotci2}), with the vanishing of $c_2$ and $c_3$ imply all the connection components are 
constants. Hence, $c_2$ and $c_3$ vanish at all times. As a result $H_2(t) = H_3(t) = 0$. The scale factors $a_2$ and $a_3$ remain constant 
in the evolution, and we choose them to be unity without any loss of generality.

Whether or not a flat Kasner metric is a solution then depends on if $\dot H_1 + H_1^2 = 0$ holds. 
To find this we notice that the Hamilton's equations in the present case  yield 
\be
\frac{\dot p_1}{p_1} = 0, ~~~ {\rm{and}} ~~~
\frac{\dot p_2}{p_2} = \frac{\dot p_3}{p_3} = \frac{1}{\gamma \lambda a_1} \sin(\bar \mu_1' c_1) ~.
\ee
Here we have used $p_2 = p_3 = 4 \pi^2 a_1$, and $a_2 = a_3 = 1$. Since $c_1$ is a constant, the Hubble rate $H_1$ can be written as  
\be\label{h12}
%H_1 = \frac{1}{2} \left(\frac{\dot p_2}{p_2} + \frac{\dot p_3}{p_3} - \frac{\dot p_1}{p_1}\right) = 
H_1 = \frac{1}{\gamma \lambda} \frac{\sin(\lambda k')}{a_1}, 
%\ee
%~~~ \mathrm{and} ~~~ 
%H_2 = H_3 = 0 ~.
\ee
where $k'$ is a constant. %we have used $p_2 = p_3 = 4 \pi^2 a_1$. %with fidciual length $l_i$ chosen to be unity.
It is easily verified that unlike the directional Hubble rate $H_1$ in $\bar \mu$ quantization, the above Hubble rate yields $\dot H_1 + H_1^2 = 0$. %Thus, a flat Kasner metric exists as a solution for the effective Hamiltonian (\ref{effham_bianchi_b}). 
Thus, $\ddot a_1 = 0$. As a result, all the spacetime curvature components vanish. 

Integrating  eq.(\ref{h12}), we get   
\be
a_1 = \alpha t ~~~ {\mathrm{where}} ~~~ \alpha = (\gamma \lambda)^{-1} \sin(\lambda k') ~.
\ee
Thus, as in the classical theory, the directional scale factor $a_1$ linearly depends on time. Since $a_2$ and $a_3$ are constants in 
the evolution, the metric of the effective spacetime 
in this loop quantization of Bianchi-I model can thus be written in the flat Kasner form with a simple redefinition of the time coordinate. 
Note that the constant $\alpha$ encodes the signature of the underlying quantum discreteness captured via $\lambda$. In the classical 
regime, $\alpha$ is unity.

Therefore, in contrast to the standard loop quantization of the Bianchi-I spacetime which does not yield a flat Kasner solution and results in 
non-vanishing of the spacetime curvature components due to underlying quantum geometry, a very different picture emerges in the 
alternative loop quantization. Even though the effective Hamiltonian is modified in a  non-trivial way, the spacetime curvature vanishes as 
in the classical theory. The effective spacetime in the $\bar \mu'$ approach is flat and the classical vacuum is unaffected by quantum 
geometry.

\section{Conclusions}
Let us summarize our main findings. Assuming the validity of the effective Hamiltonian in LQC, we studied 
the existence of flat Kasner metric in the loop quantized Bianchi-I spacetimes. At the classical level, all the components of spacetime 
curvature vanish. The classical vacuum spacetime 
can be identified as a patch of flat space. 
We considered effective Hamiltonians for two strategies to loop quantize Bianchi-I spacetime, the standard ($\bar \mu$) scheme and its 
alternative ($\bar \mu'$ scheme) studied in Sec. III and Sec. IV respectively. 
Our investigation uncovers 
 some novel features of the effective spacetime in these quantizations. For the $\bar \mu$ quantization,  we find that even in the absence of matter, Riemann curvature components are non-zero and 
the effective  spacetime is not Ricci flat. The non-vanishing of the spacetime curvature components results from the the non-perturbative quantum 
geometric modification encoded  through the minimum area of the loops used in the curvature operator. Only in the classical limit, where the 
quantum geometric effects become negligible, loop quantum effective spacetime becomes Ricci flat. Not surprisingly, the flat Kasner metric 
is not a 
solution of the effective Hamiltonian for the $\bar \mu$ quantization. 

It is intriguing that at the effective spacetime level, the loop 
quantization of a spacetime which is classically flat results in a metric which is quite non-trivial.  Unlike the classical model, the spacetime 
does not have a non-curvature singularity. 
Evolution  mimics the cases where 
a to be curvature singularity is resolved. The difference from such cases studied extensively in LQC being that there is no curvature singularity at the classical 
level which is resolved. 
Quantum gravitational effects 
become important only in a small 
neighborhood of the point where there is a classical non-curvature singularity. The Ricci tensor components and the curvature invariants remain bounded in the entire 
evolution and become negligible as the 
classical regime is approached.  On one side 
of the temporal evolution, the effective spacetime metric approaches the flat Kasner metric asymptotically in the classical limit. The Kasner 
exponents are such that one of them is unity and the other two vanish. Surprisingly, this is not the case on the other side of the 
asymptotic temporal 
evolution when the classical regime is approached. The asymptotic metric is of Minkowski spacetime with all of the Kasner exponents 
vanishing. Thus, there is a loop quantum Kasner 
transition between the asymptotic past and future in the classical regimes, but unlike of the other cases studied so far in LQC \cite{gs1}. 
Such highly asymmetric nature of the properties of the spacetime in two asymptotic regimes has parallels with recent results in the loop quantization of 
Kantowski-Sachs spacetimes \cite{djs,cs15}.

In contrast to these results, the flat Kasner metric turns out to be a solution of the effective Hamiltonian for an 
alternative loop quantization of the Bianchi-I spacetime in the absence of matter. This quantization, labeled as the $\bar \mu'$ 
quantization, does not share all the nice features of the $\bar \mu$ quantization. Physics depends on the fiducial cell, unless it is fixed 
to be a 3-torus. The expansion and shear scalar are not universally bounded and resolution of strong singularities is not guaranteed for arbitrary matter. 
In this quantization, the Riemann tensor components vanish in the effective spacetime even when the effective Hamiltonian has non-trivial quantum 
geometric modifications. Thus, we find striking qualitative differences in the 
effective spacetime of the Bianchi-I spacetime in the absence of matter for $\bar \mu$ and $\bar \mu'$ quantizations. In the $\bar \mu$ 
quantization, the spacetime which is classically vacuum is no longer vacuum  due to the presence of non-vanishing Ricci tensor components resulting from the underlying quantum geometry. 
Whereas, the classical vacuum spacetime, is vacuum in the 
effective spacetime for the $\bar \mu'$ quantization of the Bianchi-I model.

What is the status of these results at the full quantum level, beyond the effective spacetime description? One way to answer this question is by using 
the loop quantized Bianchi-I spacetimes, by considering semi-classical states peaked on the initial data as in our analysis and evolve them 
numerically using quantum Hamiltonian constraint. Since such a state has finite spread, except the peak value which classically corresponds 
to the flat Kasner solution, the state in general gets non-trivial contribution from the non-flat Kasner exponents. Given the technical 
difficulties with the quantum difference equations to perform numerical simulations in $\bar \mu$ quantization, more work is needed at the 
quantum level to gain insights on this question. The situation with $\bar \mu'$ quantization is more tractable. Here numerical 
investigations suggest that singularity is eliminated for states with a finite spread \cite{ps16b}. Such a sharply peaked state avoids a to be 
curvature singularity. Another way to gain insights on this issue is to understand whether there are any physical states in the Hilbert space  corresponding to the flat Kasner solution in $\bar \mu$ and $\bar \mu'$ quantizations. Our results suggests that for the former the answer is negative, and is positive for the latter.

Let us note some important insights which emerge when we compare our findings with existing studies in isotropic models in LQC. 
Our results reveal a non-trivial change in the nature of the vacuum in loop quantum spacetime when anisotropy is included. In the spatially 
flat isotropic model, the classical vacuum spacetime corresponds to the vacuum spacetime at the effective spacetime level in LQC for the 
improved dynamics quantization. The metric is Minkowski in the absence of matter both in the classical theory and in the effective dynamics 
at all regimes. The spacetime which at the classical level has no gravitational field, on quantization yields an effective spacetime of the 
same nature. What we find in our analysis is that this is no longer necessarily the case for the Bianchi-I spacetime. For the only known 
loop quantization of Bianchi-I model  which is free from the fiducial cell freedoms and promises generic 
resolution of strong singularities as its isotropic counterpart, i.e. the $\bar \mu$ quantization, the classical vacuum spacetime is 
non-empty at the quantum level. To restore the emptiness of vacuum one has to go beyond the 
$\bar \mu$ quantization. One way is to consider the alternative $\bar \mu'$ quantization which comes at a price: lack of genericness of 
singularity resolution and  restriction of the spatial topology to a 3-torus. Our analysis thus points to novel features of loop quantum 
Bianchi-I spacetimes which are absent in its isotropized form. Our results show that the 
simplicity of the spatially flat isotropic model in LQC masks some of the subtle features of the loop quantum spacetimes with more gravitational degrees of freedom.

\section*{Acknowledgments}
We thank Brajesh Gupt and Jorge Pullin for discussions during initial stages of this investigation, and Sahil Saini for comments and discussions.
This work is supported by NSF grants PHY-1403943 and PHY-1454832.

%\begin{thebibliography}{000} %for 3 digits
%\begin{thebibliography}{00}  %for 2 digits

\end{document}